\documentclass{PoS}

\usepackage{amsmath}

\usepackage[square, comma, sort&compress, numbers]{natbib}
\newcommand{\beq}{\begin{equation}} 
\newcommand{\eeq}{\end{equation}} 
\newcommand{\bea}{\begin{eqnarray}} 
\newcommand{\eea}{\end{eqnarray}} 
\newcommand{\nn}{\nonumber} 
\newcommand{\refeq}[1]{(\ref{#1})}

\title{The Inert Doublet Model and its Phenomenology}

\ShortTitle{The Inert Doublet Model and its Phenomenology}

\author{\speaker{Michael Gustafsson}
 \\
        Dipartimento di Fisica Galileo Galilei, Via Marzolo 8, 35131 Padova - Italy\\
        E-mail: \email{michael.gustafsson@pd.infn.it}}

\abstract{The single Higgs doublet in the standard model (SM) may be the simplest way of introducing electroweak symmetry breaking, but SM extensions with more scalar doublets are not excluded. A special case of the two Higgs doublet models is the inert doublet model -- a  minimalistic version with interesting phenomenology. These proceedings reviews the inert doublet model's theoretical setup, constraints, collider prospects and its dark matter phenomenology.}

\FullConference{Third International Workshop on Prospects for Charged Higgs Discovery at Colliders - CHARGED2010,\\
		September 27-30, 2010\\
		Uppsala Sweden}

\begin{document}
\section{Introduction} 
Despite its simplicity, the Inert Doublet Model (IDM) can offer a variety of phenomenological and theoretically appealing properties while being reliably perturbatively-calculable, consistent with current data and still offering a wealth of signals reachable with current and upcoming experiments. Results within IDM also show similarities to other models that have a modified scalar sector of the standard model (SM), and the IDM could therefore serve as a good archetype model. 

Historically the IDM was introduced in the 1970s \cite{Deshpande:1977rw}, then reached new attention when the model was shown to be able to ameliorate the `LEP paradox' \cite{Barbieri:2000gf,Barbieri:2006dq,Casas:2006bd}, generate light neutrino masses via an one-loop radiative see-saw mechanism \cite{Ma:2006km} and leptogenesis \cite{Ma:2006fn} by including TeV scale right-handed neutrinos, have electroweak symmetry breaking induced by loop effects \cite{Hambye:2007vf}, achieve grand unification by putting the inert doublet in a {\bf 5} representation of a discrete symmetry group  \cite{Lisanti:2007ec}, and, the main focus in this review, provide a thermally produced dark matter (DM) candidate \cite{Barbieri:2006dq,LopezHonorez:2006gr,Gustafsson:2007pc,Lundstrom:2008ai,Hambye:2009pw,Dolle:2009fn,Honorez:2010re,LopezHonorez:2010tb}.

\section{The theory}
The IDM consists of the SM, including its Higgs doublet $H_1 = $${0}\choose{v + h/\sqrt{2}}$, and an additional Lorentz scalar SU(2) doublet $H_2 =$${H^+}\choose{(H^0+A^0)/\sqrt{2}}$. The added, so called, inert doublet has a standard kinetic gauge term $[D^\mu H_2]^\dagger [D_\mu H_2]$, and, what singles out the IDM from more general two Higgs doublet model, its potential has a $\mathbb Z_2$ symmetry that is unbroken by the vacuum state. Specifically the Lagrangian is imposed to be \emph{invariant} under the $\mathbb Z_2$ \emph{parity transformation} where $H_2 \rightarrow - H_2$ and all the other (SM) fields are even, $\psi_{\rm SM} \rightarrow +\psi_{\rm SM}$. Such an unbroken discrete symmetry guarantees the absence of Yukawa couplings between fermions and the inert doublet $H_2$ (hence its prefix \emph{inert}) and therefore that no tree-level neutral flavor changing currents appear. 
%
The most general renormalizable CP conserving potential for the $H_2$ field is then
\bea \label{eq:potential}
\mbox{ }\hspace{-0.5cm} 
V&=& \mu_1^2 \vert H_1\vert^2 + \mu_2^2 \vert H_2\vert^2+ \lambda_1 \vert H_1\vert^4 + \lambda_2 \vert H_2\vert^4
\cr 
&&\quad 
+ \lambda_3 \vert H_1\vert^2 \vert H_2 \vert^2 + \lambda_4 \vert H_1^\dagger H_2\vert^2 + 
{\lambda_5} Re[(H_1^\dagger H_2)^2],\;
\eea
where $\mu_{1,2}^2$ and $\lambda_{1-5}$ are real parameters. Except for the `SM Higgs', $h$, four new physical scalar particle states are present: two charged $H^{\pm}$, and two neutral $H^0$ and $A^0$ (with opposite CP-parities).  After standard electroweak symmetry breaking, the masses of the scalar particles are given by:
\bea \label{eq:masses}
  m_h^2 &= & -2\mu_1^2 = 4 \lambda_1 v^2,\quad\cr
  m_{H^\pm}^2 & =& \mu_2^2 +  \lambda_3 v^2,\quad \cr
  m_{A^0}^2   &=& \mu_2^2 + \lambda_A v^2  \quad\quad\quad ({\rm where \;} \lambda_A = \lambda_3 + \lambda_4 - \lambda_5),\cr
  m_{H^0}^2   &=& \mu_2^2 + \lambda_L v^2 \quad\quad\quad ({\rm where \;} \lambda_L = \lambda_3 + \lambda_4 + \lambda_5),
\eea
where $v \approx175$ GeV is the vacuum expectation value (vev) of the Higgs field $H_1$ (the only field responsible for the symmetry breaking). The model has 6 independent parameters in the scalar sector (if we fix the $\mu_1$ to $\lambda_1$ ratio by the  SM vev).
If a field theory model has a massive, stable, chromodynamic and electromagnetic uncharged particle it has the potential to provide a good DM candidate -- a weakly interacting massive particle (WIMP). The IDM can provide such a DM candidate. In the following we take $H^0$ as the lightest inert particle (LIP), and hence our WIMP DM candidate (taking $A^0$ as the lightest inert state would be an equivalent choice).

\section{Constraints}
A number of theoretical, observational (to be discussed in sections 4 and 5) and experimental constraints can be imposed on the model parameters ($\mu_{1,2}$ and $\lambda_{1,2,3,4,5}$) appearing in the potential \refeq{eq:potential}, or equivalently on, say, the four masses in \refeq{eq:masses} plus the couplings $\lambda_2$ and $\mu_2^2$. 
%
%
\begin{itemize}
\item[] \hspace{-9mm}{\bf Vacuum stability:}
With the squared masses in \refeq{eq:masses} positive (incl.\ $m_{H^\pm}^2\!>m_\mathrm{LIP}^2$), the potential \refeq{eq:potential} is (a) bounded from below and (b) a global minimum that preserves the $Z_2$-symmetry iff\footnote{Possible evolutions of the vacuum state during cosmological cooling has been studied within the IDM in ref.~\cite{Ginzburg:2010wa}.}:
\bea
\hspace{-10mm}
    \mathrm{a)} \; \lambda_{1,2} > 0 \;\; \mathrm{;}\;\; 
    \lambda_{3},\;\lambda_{3}+\lambda_{4}-|\lambda_{5}| > -2 \sqrt{\lambda_{1}\lambda_{2}}
    \quad \mathrm{and} \quad
    \mathrm{b)} \; \frac{\mu_1^2}{\sqrt{\lambda_1}} < \frac{\mu_2^2}{\sqrt{\lambda_2}}.
\eea

%
\item[] \hspace{-9mm}{\bf Perturbativity:}
In perturbation theory calculations, the expansion parameters ({\it i.e.}\ the relevant couplings) should not be too large. As a rule of thumb one might adopt $|\lambda_i| \lesssim$ few
(or $< 4\pi$). In \cite{Barbieri:2006dq} \emph{sufficient} conditions for keeping the runnings of scalar couplings under control up to the  TeV scale was derived:
$\lambda_3^2 + (\lambda_3+\lambda_4)^2 + \lambda_5^2 \lesssim 12 \lambda_1^2$, $\lambda_2\lesssim1$ and $m_h\lesssim700$ GeV \cite{Djouadi:2005gi}.

\item[] \hspace{-9mm}{\bf Colliders searches:}
Precision measurement results by LEP-I  exclude the possibility that massive SM gauge bosons decay into inert particles, which consequently requires that \cite{Gustafsson:2007pc,Cao:2007rm}:
{$m_{H^\pm}+m_{H^0,A^0} \gtrsim m_{W^\pm}$},\;
{$m_{H^0}+m_{A^0},\; 2 m_{H^\pm} \gtrsim m_{Z}$}.
No dedicated analysis of LEP-II data to search for the IDM has been carried out. However, by comparing with supersymmetry studies one has indirectly derived that: $m_{H^\pm} \gtrsim 70 - 90$ GeV  \cite{Pierce:2007ut}, and excluded masses in the ($m_{H^0}$, $m_{A^0}$) plane 
fullfilling \cite{Lundstrom:2008ai}
\beq
m_{H^0} \lesssim 80\; \mathrm{GeV} \;\;\wedge\;\; m_{A^0} \lesssim 100\, \mathrm{GeV} \;\;\wedge\;\; m_{A^0}-m_{H^0}  \gtrsim 8\mathrm{GeV}.
\eeq
Direct collider searches for the SM Higgs at LEP put a lower Higgs mass bound of 114.4~GeV and the Tevatron 
exclude an additional mass region from $\sim$158 to 175 GeV. These SM Higgs searches do not directly translate into mass limits on the inert scalars; as their odd $\mathbb Z_2$ parity leaves $H^0$ invisible for colliders and forces inert particles to always be produced in pairs. Inert states do however modify Higgs decay widths, and in \cite{Cao:2007rm} they showed that modified `SM Higgs' decay channels 
can {\it e.g.}\ modify the LEP bound to read $m_h \gtrsim$ 105 GeV. 

\item[] \hspace{-9mm}{\bf Electroweak precision tests (EWPT):}
Loop-level induced effects indirectly limit the SM Higgs boson to have a mass below about 160 GeV \cite{LEPHiggsLimit}. This conclusion may however change if new particles contribute to cancel SM loop-effects. This is what can happen within the IDM, and a heavy `SM Higgs' up to $\sim$600 GeV is possible without fine-tuning \cite{Barbieri:2006dq}\footnote{This improved naturalness has however been disputed by the authors of reference \cite{Casas:2006bd}.}. Interestingly, this would ameliorate the so called `LEP paradox'\cite{Barbieri:2000gf} 
-- the apparent paradox that (LEP) data indicates both a light SM Higgs $\lesssim 160$ GeV and no new generic physics before multi TeV energies; which in turn means that the SM must be fine-tuned in order to cancel the large radiative corrections to the Higgs mass that is expected to diverge up to the scale where new divergence canceling physics (such as {\it e.g.} supersymmetry) can appear.

The EWPT observables are commonly parametrized into the so called Peskin-Takeushi parameters $S$, $T$ and $U$. A heavy SM Higgs boson gives rise to a too small value of the $T$ parameter. Similarly, the IDM makes an important contribution only to $T$. To first order the shift in $T$, from a SM reference point, depends on the scalar masses as \cite{Barbieri:2006dq}:
\bea
\Delta T_{\mathrm SM} \approx - \frac{3}{8 \pi \cos^2{\theta_W}} \ln(\frac{m_h}{m_Z}) &,&\quad
\Delta T_{\mathrm IDM} \approx
\frac{F(m_{H^\pm},m_{A^0})+F(m_{H^\pm},m_{H^0}) - F(m_{A^0},m_{H^0})}{32 \pi^2 \alpha v^2},
\nn\\
&&\quad F(m_1,m_2) = \frac{m_1^2+m_2^2}{2} - \frac{2 \, m_1^2 m_2^2}{m_1^2-m_2^2} \ln\frac{m_1}{m_2},
\eea
where $\theta_W$ is the Weinberg angle, $\alpha$ the fine structure constant and $m_Z$ the $Z$ boson mass.  
Experimentally $\Delta T = \Delta T_{\mathrm SM}  + \Delta T_{\mathrm IDM}$ is bounded to be roughly in the range $0.1 - 0.3$.

\item[] \hspace{-9mm}{\bf Naturalness:}
If a model aims to ameliorate the `LEP paradox' by enabling a heavier SM Higgs boson, and by that reduce the fine-tuning needed in the SM, then neither any of the other parameters of the model should  need fine-tuning. A set of technical naturalness constraints on IDM,  {\it i.e.}\ less fine tuning than 1 part in $D>1$ up to the scale $\Lambda_{\mathrm cut}\sim TeV$, is found in \cite{Barbieri:2006dq}:
\bea
 \hspace{-1.5cm}  m_h^2  > 
  \left\{   |\alpha_{h,g,t}|       ,\;\frac{|2 \lambda_3+\lambda_4|}{8 \pi^2}   \right\}     \times\frac{\Lambda_{\mathrm cut}^2}{D}  
\quad\mathrm{and}\quad
  |\mu_2^2|  >  
  \left\{ \frac{|\alpha_g|}{2}  ,\;     \frac{3 |\lambda_2|}{8\pi^2} , \;   \frac{|2 \lambda_3 + \lambda_4|}{16 \pi^2}  \right\}  \times \frac{\Lambda_{\mathrm cut}^2}{D},
\eea
where $\alpha_{h,g,t} =  -\left\{3 m_h^2 ,\;  6 m_W^2+3 m_Z^2 ,\;-12 m_t^2 \right\} /(16 \pi^2 v^2)$.
\end{itemize}

\section{Dark matter}
Observationally cold DM make up $\Omega_{\rm CDM} = (22\pm2)\,\%$ of our ($\Lambda$CDM) universe  \cite{Larson:2010gs}, and the abundance of thermally produced DM can in such a scenario be calculated by solving the Boltzmann equation, which describes the time evolution of the number density of WIMPs. A guideline is that $\Omega_{\rm WIMP}  \approx \frac{ 6\times 10^{-27}  \mathrm{cm}^3/\mathrm{s}}{\langle \sigma v \rangle}$, where $\langle \sigma v \rangle$ is the effective WIMP annihilation cross-section\,$\times$\,velocity at freeze-out. 
It turns out that the $H^0$ particle can provide a good DM candidate only in restricted $m_{H^0}$ mass regions \cite{Barbieri:2006dq,LopezHonorez:2006gr,Gustafsson:2007pc,Lundstrom:2008ai,Hambye:2009pw,Dolle:2009fn,Honorez:2010re,LopezHonorez:2010tb}. To understand this let us first formulate the interactions $H^0$ have:
\smallskip
\\
\mbox{ }\, (a)  The Lagrangian's kinetic term has standard gauge-couplings, that give rise to the processes: \\
\indent \; $H^0 H^0 \rightarrow W^+W^-,ZZ$ \;and co-annihilations\; $H^0 A^0 \xrightarrow{Z} SM$ and $H^0 H^\pm \xrightarrow{W^{\pm}} SM$.
\\
\mbox{ }\, (b) The scalar potential \refeq{eq:potential} gives couplings to the SM Higgs boson and the processes:\\
\indent \; $H^0 H^0 \rightarrow h h $ \; and\;  $H^0 H^0 \xrightarrow{h} SM$.

\bigskip
\noindent
Illustrated in Fig.\,\ref{fig:1}, there are three distinct IDM DM mass regions:
\begin{figure}[t!] 
\centering{\resizebox{0.99\linewidth}{!}{\includegraphics{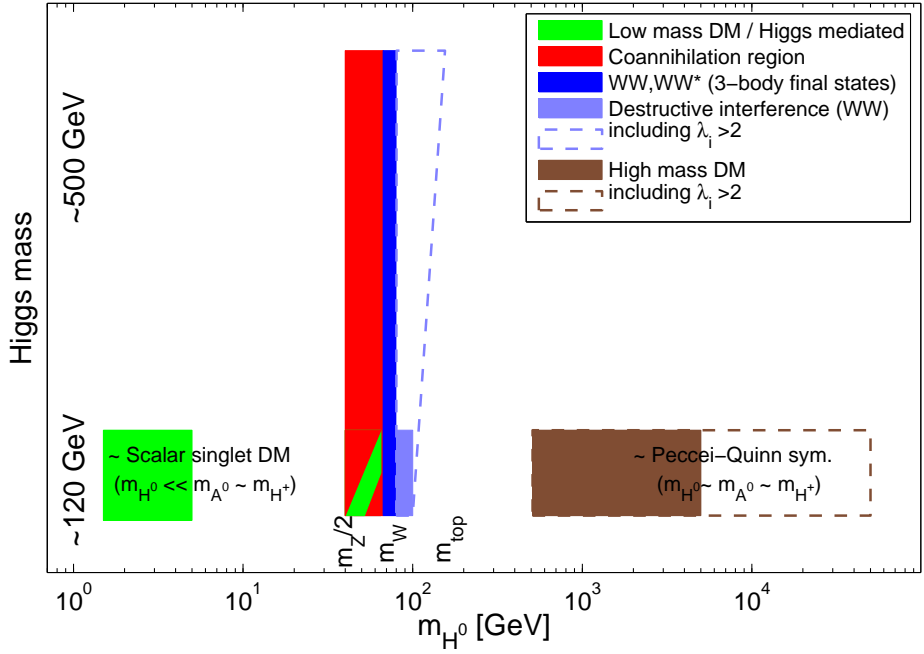}}}
  \caption{Schematic figure of allowed $H^0$ dark matter (DM) mass regions in the inert doublet model (IDM). The intermediate mass region 40 GeV $\lesssim m_{H^0} \lesssim 160$ GeV is particular for IDM, whereas the low mass regime 1 GeV $\lesssim m_{H^0} \lesssim 5$ GeV have clear  similarities to singlet scalar DM models \cite{Burgess:2000yq,Andreas:2010dz} and the high mass regime 500~GeV $\lesssim m_{H^0} \lesssim 50$ TeV show similar phenomenology as other electroweak minimal DM models \cite{Cirelli:2005uq}. When the scalar couplings $\lambda_i$ are allowed to be large ($\gtrsim 2$) an enlarged parameter space is possible while still compatible with DM relic density and experimental constraints. In the intermediate mass range there are four different annihilation processes to consider: annihilation via a Higgs boson, coannihilation with~{\it e.g.}~$A^0$
, annihilation into 3-body final states ($WW^* \rightarrow W f f'$), and annihilations into real gauge bosons ($WW$ and $ZZ$) with suppressed amplitudes duo to destructive interference among contributing Feynman diagrams.}
 \label{fig:1} 
\end{figure} 

\begin{itemize}
\item[] \hspace{-9mm}{\bf Low mass} (1 GeV $\lesssim m_{H^0} \lesssim 5$ GeV):  For low $H^0$ masses there must be a large mass gap up to both $A^0$ and $H^\pm$ to satisfy collider constraints. The annihilations of $H^0$  can therefore only be through the second process in (b) -- annihilation via the Higgs boson into massive fermions. This mimics singlet scalar models \cite{Burgess:2000yq,Andreas:2010dz}. To have correct cross section ($\propto \bar{\sigma} \equiv \lambda_L^2 m_f^2/m_h^4$) at freeze-out the coupling between $H^0$ and the Higgs boson ($\propto\lambda_L$) must be fairly large. At $m_{H^0}\lesssim 1$\,GeV the IDM DM can no longer be compatible with the perturbativity bounds; as a stable vacuum with no vev for $H_2$ requires $\lambda_2 \gtrsim \lambda_L^2 v^2/m_h^2 = \bar{\sigma} m_h^2 v^2/m_f^2$ (similar to eq.~(24) in \cite{Andreas:2009hj}). 
%
%
On the other mass end, when $m_{H^0}\gtrsim 5$\,GeV up to around 40 GeV the IDM DM would overshoot recent \cite{Angle:2011th} direct DM detection bounds \cite{Arina:2009um}. By imposing also gamma-ray, radio, CMB and antiproton constraints, discussed in the next section, this whole low mass IDM DM region might already be considered to be challenged.
\item[] \hspace{-9mm}{\bf Intermediate mass} (40 GeV $\lesssim m_{H^0} \lesssim 160$ GeV):
The correct DM relic density can in this mass regime be achieved via different processes: 
\\
\mbox{ }\;\, \indent (i) $H^0$ pair annihilations via s-channel Higgs boson into fermions (mainly $b \bar b$ and $\tau^+\tau^-$), 
\\
\mbox{ }\, \indent (ii)  coannihilations $H^0 A^0 \xrightarrow{Z} SM$ (and other coannihilation processes),
\\
 \mbox{ } \indent (iii) annihilations into massive gauge bosons $H^0H^0 \rightarrow W^+W^-,\,ZZ$, and potentially into $hh$,
\\
 \mbox{ } \indent (iv) annihilations into 3-body final states, $H^0H^0 \rightarrow WW^*\rightarrow W ff'$ (where $f$ is a fermion).\\
For a heavy Higgs ($\gtrsim 300$ GeV) process (i) is not sufficient enough unless $\lambda_L$ is tuned to be large ($\gtrsim 1$) or   it takes place close to the Higgs resonance. For (iii) to work cancellation between Feynman diagrams of type (a) and (b) has to occur since the `strong' gauge couplings otherwise would deplete the $H^0$ WIMP density too much \cite{LopezHonorez:2010tb}.  For (iii), a sufficient mass hierarchy between $H^0$ and $H^\pm$ (corresponding to $\lambda_3 \gtrsim$ 1) is also typically needed to suppress the t/u-channel diagrams contributing to $H^0$ annihilation into $W^+W^-$.
As the annihilation cross sections into massive gauge bosons can be large, one can also understand why the 3-body process (iv) can be important in the vicinity below the $WW$ kinematical threshold ($m_{H^0} \lesssim m_W \approx 80.4$\,GeV)  \cite{Honorez:2010re}. 
\item[] \hspace{-9mm}{\bf High mass} (500 GeV $\lesssim m_{H^0} \lesssim 50$ TeV):
For $H^0$ masses above the top quark mass $m_t$, all annihilation channels are open and annihilation cross sections are large. At the highest $H^0$ masses the cross sections should however eventually drop at least as $1/m_{H^0}^2$ due to the unitarity condition \cite{Griest:1989wd,LopezHonorez:2006gr}, and a high DM mass region should be present. The right relic density can be achieved at $m_{H^0} \approx 534$ GeV (which happen when all the inert scalars are mass degenerated)~\cite{Hambye:2009pw}. To keep correct effective cross section also for increased $m_{H^0}$ the scalar self couplings $\lambda_i$ can be tuned properly, but at a few TeV some of them approach 1, and at  $m_{H^0} \approx 60$ ~TeV it is not possible to keep all the interaction couplings ($\lambda_L$, $\lambda_A$, $\lambda_3$) below $4\pi$. In the high mass DM regime, the mass splitting is always small, which enforces a negligible contribution to $\Delta T_{\mathrm IDM}$. EWPT therefore require a light SM Higgs boson $\lesssim 160$~GeV~\cite{Hambye:2009pw}. In this mass region the IDM has phenomenology similar to other minimal DM models ({\it cf.\ e.g.}\ \cite{Cirelli:2005uq}).
\end{itemize}

\section{Phenomenology}
There is a diversity of indirect, direct and particle collider signals an IDM could give rise to. Indirect DM signals arise when $H^0$ particles pair annihilate and their rest mass energy is injected into SM particles, with well predictable energy spectra, that directly or indirectly give signals to search for with telescopes. Astrophysical uncertainties still make some of these signal strengths
hard to predict and/or separate from backgrounds accurately. Direct DM searches, looking for recoil events induced by WIMPs scattering in the detectors, are also subject to astrophysical as well as nuclear interaction uncertainties. None the less, these type of signals can lead to both robust constraints on models and  give prospects for identifiable DM signals.
%
%

\bigskip
\noindent{\bf Gamma rays, X-rays and radio waves:}\\
In the {\bf low mass} region, the annihilation mechanism today is the same as at freeze-out -- Higgs mediated annihilations into mainly quarks and taus. A continuum of gamma-ray energies are produced mainly due to the quarks hadronization processes, where pions are created and subsequently decay into photons. The Fermi-LAT gamma-ray telescope sets gamma-ray flux limits that are in tension with the expected IDM signals in this mass regime \cite{Andreas:2010dz,Arina:2010rb,Fermi}.  Similarly, annihilations inject electrons and positrons, which in a radiation and magnetic-field environment, lead to additional gamma ray, \mbox{X-ray} and radio signals due to {\it e.g.}\ Bremsstralung, inverse Compton and synchrotron emission. 
Radio signal constraints from the galactic center region can exclude (under reasonable DM density and magnetic field assumptions) the lowest mass region \cite{Boehm:2010kg}. Also the constraints from no detected distortions of the CMB due to WIMP annihilations challenge the low mass IDM region \cite{Hutsi:2011vx}. 
In the {\bf intermediate mass} range, a particularly clear astrophysical DM signal in gamma-rays is possible: A monochromatic photon line produced at one-loop level, via {\it e.g.}\ virtual massive gauge bosons running in a Feynman loop \cite{Gustafsson:2007pc}. 
In the mass range around $m_{H^0} \sim 50$-60 GeV, where 
coannihilation could have dominated at freeze-out and annihilations into \mbox{3-body} final states $H^0H^0 \rightarrow WW^* \rightarrow W f f'$ is not very strong, these $H^0H^0 \rightarrow \gamma\gamma$, and to less extent $H^0H^0 \rightarrow \gamma Z$, annihilations directly into monochromatic gamma-lines can have fairly large cross sections and branching ratios of several percent \cite{Gustafsson:2007pc}. An observed multi GeV gamma-ray line in the sky would be a striking DM signal, but so far Femi-LAT have not observed any line signal in this energy range \cite{Abdo:2010nc,Vertongen:2011mu}.  At the moment this only excludes IDM scenarios with strong gamma-ray lines in combination with a strong astrophysical signal-enhancement factor $\gtrsim 10$ (such an enhancement of the signal, compared to a vanilla Navarro-Frenk-White distribution of the DM in our Galaxy, could potentially come from a pronounced DM accretion around the super massive black hole in the Galactic center or to numerous dense substructures in the Galactic halo) \cite{Gustafsson:2007pc}.
At higher $H^0$ masses, including the {\bf high mass regime}, pair production of the massive gauge bosons is kinematically accessible. A large amount of  SM particles are injected at each annihilation, but as the number density of DM particles drops for heavier DM particles a weaker signal is still expected. Signals fall below observational limits \cite{Bertone:2008xr} unless  {\it e.g.}\ a significant boost $\gtrsim100$ is present due to a possible Sommerfeld enhancement \cite{Cirelli:2007xd} for $m_{H^0}\sim10$~TeV.

\medskip
\noindent{\bf Neutrinos:}\\
\noindent Massive celestial bodies, like the Sun and the Earth, moving through the dark halo can gravitationally trap DM particles in their cores. Neutrinos from WIMP annihilations in these high density cores can escape and be searched for by neutrino telescopes. 
In the {\bf low mass} IDM regime, neutrino signals from the Sun are more promising than from the Earth \cite{Agrawal:2008xz,Andreas:2009hj}. The Super-Kamiokande experiment currently put the strongest neutrino limits and challenges the $m_{H^0}\sim3-4$~GeV mass range \cite{Andreas:2009hj,Kappl:2011kz}. 
In the {\bf intermediate mass} range the signal from the Earth can be larger due to the possibility of resonance effects, with {\it e.g.}\ iron, producing larger $H^0$ capture cross-sections. The authors of \cite{Andreas:2009hj}  did however not find any IDM DM model in the intermediate mass region that can be reached by the 1~km$^3$ IceCube detector and still be compatible with direct detection exclusion limits.
In the IDM's {\bf high mass} regime the capture cross sections on nuclei ($\propto1/m_{H^0}^2$ ) are too small and the most promising neutrino signal comes instead from the Galactic center. In the vanilla scenario a neutrino signal is not visible, and for ANTARES to detect a signal an exceptionally high DM density in the Galactic center and/or a  large Sommerfeld enhancement boosting the signal $\sim10^3$ would be needed to give a detectable signal \cite{Andreas:2009hj}.

\medskip
\noindent{\bf Positron and Antiprotons:}\\
Cosmic-ray signals from the IDM were studied in \cite{Nezri:2009jd}. The IDM is not expected to have strong positron signals due the absence of direct coupling to fermions. 
In the {\bf low mass} region the positron signal is a factor $\sim 10$ below observations \cite{Nezri:2009jd}, whereas the predicted signal in antiproton fluxes are comparable to, or may overshoot, observed upper flux limits \cite{Nezri:2009jd,Lavalle:2010yw}.
In the {\bf intermediate mass} region the positron flux is also within observational limits, but could become comparable to positron fraction data observed below $\sim 10$ GeV within the setup studied in \cite{Nezri:2009jd}. Cosmic-ray propagation have intrinsic uncertainties though and, at these low energies also solar modulation effects are important. The antiproton fluxes could also in this mass range be interesting, but a flux enhancement by more than ten compared to vanilla astrophysical assumptions are needed to indicate a mismatch with current data in the antiproton to proton ratios. 
In the {\bf high mass regime} the model is similar to the minimal DM scenario; early proposed to fit the PAMELA positron fraction data \cite{Cirelli:2008jk}. Annihilation into massive gauge bosons can give hard positron and antiproton energy spectra, but due to the large DM particle mass the signal is low (as the DM number density drops as $\propto 1/m_{\mathrm WIMP}$). For example, to fit the PAMELA positron data an annihilation enhancement of the order of $10^4$ would be needed for a 10 TeV $m_{H^0}$. With recent antiproton data \cite{Adriani:2011cu} and photon constraints there is a significant tension between such a strong positron signal from IDM DM and other observational data (see {\it e.g.}\ \cite{Meade:2009iu}).

\bigskip
\noindent{\bf Direct detection:}\\
At tree level, there are two types of interactions by which $H^0$ particles can deposit kinetic energy to an atom nucleus in direct detection experiment: $H^0 q\xrightarrow{Z}A^0 q$ and $H^0 q\xrightarrow{h} H^0 q$, where $q$ is a quark in a nucleon in the nucleus \cite{Barbieri:2006dq,Majumdar:2006nt,LopezHonorez:2006gr}. The former process, with a $Z$ exchange, is very strong and forbidden by current experimental limits. However, if the mass splitting $\Delta = m_{A^0}-m_{H^0}$ is more than a few 100 keV this process is kinematically forbidden, as the typical kinetic energy transfer in a $H^0$ scatter would then not be enough to excite $H^0$ into $A^0$. At the same time, it has been shown that for {\bf high mass} $H^0$ this inelastic scattering can be tuned to give a signal that fits the long standing/controversial DAMA/LIBRA \cite{Bernabei:2008yi}  signal when $\Delta \sim100$\,keV and the $H^0$ mass is a few TeV \cite{Arina:2009um}.\footnote{Such a small mass splitting might be protected by the exact Peccei-Quinn symmetry emerging when $\Delta=0$.}
However, recent results presented by {\it e.g.}\ the CRESST \cite{CRESST:2011} and the XENON100 \cite{Aprile:2011ts} collaborations exclude this possibility.
In the {\bf intermediate mass} range direct detection experiments set relevant limits on the IDM parameter space. As already mentioned, in the $H^0$ mass range $\sim 5$ GeV to $40$ GeV IDM DM is already excluded from the Higgs-mediated elastic scattering rates they would produced in direct detection experiment. Note that due to LEP constraints coannihilations cannot occur at freeze-out in this mass range, and the Higgs mediated coupling is thus fixed by the relic density constraint.  For $m_{H^0} \gtrsim 40$~GeV, coannhilations (process ii) at freeze-out and being closer to the $W^+W^-$ production threshold (process iii-iv) allows nucleon interactions to be much lower today and compatible with direct detection DM limits. For the IDM DM region $ 80\;\mathrm{GeV} \gtrsim m_{H^0} \gtrsim 160\;\mathrm{GeV}$, the recent XENON100 result \cite{Xenon100:2011hi} is now close to eliminating this whole region presented in \cite{LopezHonorez:2010tb}.
In the {\bf low mass} IDM region the model can mimic singlet scalar singlet DM, that interact only via the Higgs portal. This type of light scalar DM models has been shown to automatically fit into disputed DAMA/Libra \cite{Bernabei:2008yi} and CoGeNT \cite{CoGeNT} preferred signal regions when the cross section is fixed by requiring correct amount of thermally produced DM ({\it e.g.} \cite{Andreas:2010dz}). This attractive possibility is however challenged by 
the CDMS data \cite{Savage:2010tg} and the reanalysis of XENON10 data \cite{Xenon100:2011hi}.

\bigskip
\noindent{\bf Hadron Colliders:}\\
The potential for detecting IDM signatures at hadron colliders was first explored in \cite{Barbieri:2006dq,Cao:2007rm}. As the IDM does not have a QCD sector, inert scalars can only be produced via electroweak interactions and relatively small production cross-sections are therefore expected. The {\bf high mass} regime of the IDM is beyond reach for LHC, but for very {\bf low mass} the couplings to quarks could potentially be constrained by the Tevatron's limits on mono-jet events \cite{Bai:2010hh,Goodman:2010ku}. However, for real scalars, like $H^0$, these limits translate into weak bounds. 
The discovery potential for a set of IDM benchmark models with $H^0$ in its {\bf intermediate mass} range has also been studied. All the studied models showed more than $3\sigma$-significance detection potential in the dilepton channel at 100 fb$^{-1}$ and a center-of-mass energy of 14 TeV \cite{Dolle:2009ft}. Some models even showed a $3\sigma$ discovery potential already at 10 fb$^{-1}$. For the trilepton channel, a couple of benchmark models showed $3\sigma$ detection potential at a luminosity of 100~fb$^{-1}$~\cite{Miao:2010rg}, but all those had a SM Higgs boson lighter than 150 GeV. The complementary, almost background-free, multilepton signal with $\geq 4$ leptons seem difficult to discover at early stages of the LHC runs \cite{x}. Note that most of the LHC studied IDM benchmark models, except those with substantial coannihilation or $WW$ annihilation at freeze-out, are now in conflict with the recent, {\it e.g.}\ XENON100 \cite{Xenon100:2011hi}, direct detection bounds. Even if one of these LHC channels would show a clear signal of beyond SM physics they will not reveal the exact nature of the new physics. A combination of signals in many complementary channels at LHC and other probes, like the ones mentioned above, are most likely needed to pin point a DM candidate. 

\section{Summary}
The IDM provides a scenario with a rich phenomenology despite its simplicity. Existing experimental and observational data provide already very crucial limits on the parameter space, but with properly chosen parameters the lightest inert state gives a good DM candidate with reachable astrophysical signals: a striking gamma-ray line, DM direct detector events (and, although in tension with the other data, as a model providing a fit to controversial direct detection `DM signal' data), charged cosmic-ray fluxes in antiprotons and to a less extent in positrons, and finally neutrino events that in some regions provide relevant complementary constraints on the IDM. LHC signals in the lepton channels, detection of a heavy `SM Higgs' boson and/or deviations in the decay width from a pure SM Higgs boson could eventually give complementary hints of an IDM like scenario.

\bigskip
\bigskip
\noindent{\bf Acknowledgments.}
I would like to thank the organizers for inviting me to the c{\Large H}\raisebox{7pt}{\scriptsize $\pm$}\!\!\!\!arged 2010 workshop and for the research support from the Fondazione Cariparo Excellence Grant `LHC and Cosmology'.


\end{document}